# Phenomena from intersecting branes in six-dimensional Anti-de Sitter spacetime

B F Riley

AMEC NNC Limited, Solutions Business, The Renaissance Centre, 601 Faraday Street, Birchwood Park, Birchwood, Warrington WA3 6GN, UK.

E-mail: bernard.riley@amecnnc.com

**Abstract**
We relate four-dimensional mass parameters to the positions of AdS 4-brane intersections in six-dimensional AdS spacetime. The 4-branes wrap 1-cycles on the rectangular toroidal orbifold $T^2/(Z_2)^3$, are situated upon and parallel to the fixed lines, and intersect orthogonally on four-dimensional Minkowski junctions. Particles are arranged upon the line equidistant from the two extra dimensions. In four dimensions, the particles occupy a geometric sequence of mass levels that descends from the Higgs field vacuum expectation value. Scales that we identify with the three Standard Model families derive from the geometry of the orbifold and stand in precise relationship to the GUT scale of the Minimal Supersymmetric Standard Model.

## 1. Introduction

We begin by briefly reviewing the literature, in particular the brane-world models, that have informed this work. Brane-world models exploit the geometry of extra dimensions to address the hierarchy problem of the gulf in size between the Planck scale, ~$10^{19}$ GeV, and the weak scale, ~$10^3$ GeV [1-4]. In the Randall and Sundrum RS I model [3], an extra spatial dimension is compactified on an $S^1/Z_2$ orbifold and the Standard Model particles and forces are confined to a negative-tension 3-brane, the 'visible' brane, separated from a positive-tension 3-brane, the Planck or 'hidden' brane, by a slice of five-dimensional Anti-de Sitter spacetime (AdS$_5$). Einstein's equations are solved to find the metric

$$ds^2 = e^{-2k|y|}\eta_{\mu\nu}dx^\mu dx^\nu + dy^2, \qquad (1)$$

where $\eta_{\mu\nu}$ is the Minkowski metric of four-dimensional spacetime, $k$ is the AdS curvature and $y$ is the coordinate of the extra dimension. The exponential factor is the source of the hierarchy between the Planck and weak scales; it results from the small overlap in the fifth dimension of the graviton wave function with our 'visible' 3-brane. In the RS II model [4], the negative tension brane is moved off to infinity; the visible brane is at $y = 0$. On the basis of the Anti-de Sitter/Conformal Field Theory (AdS/CFT) conjecture [5, 6], the coordinate $y$ can be thought of as parametrizing the four-dimensional scale [7]. Two field excitations at scales related by a transformation factor $e^{-\lambda}$ in four-dimensional spacetime correspond to two excitations centred on positions related by a translation $y \to y + \lambda/k$ in the fifth dimension. We shall make use of this correspondence when relating four-dimensional scales, and particle masses, to positions within an extra dimension. The AdS/CFT conjecture relates compactifications on AdS spacetimes of ten-dimensional string theories, and eleven-dimensional M-theory [8, 9], to conformal field theories in various dimensions.

Lykken and Randall have combined the results of RS I and II to address the hierarchy problem in an infinite extra dimension with positive tension branes only [10]. Gravity is localized on the Planck brane and we live on a separate brane in the fifth dimension, upon which mass scales are exponentially suppressed. Oda has shown how gravity can be trapped on multiple D3-branes, with positive tension, situated at intervals along the extra dimension [11]. It was later shown that fields of all spins (≤2) can be localized on positive tension branes [12].

Supersymmetric versions of the Standard Model have been constructed with an extra dimension compactified on $S^1/Z_2$ [13] or $S^1/(Z_2 \times Z_2')$ [14, 15]. Orbifold compactifications are used to break the symmetries of the higher dimensional theory and produce a four-dimensional effective theory [16, 17].

In an extension of the Randall and Sundrum RS II model [4], by Arkani-Hamed, Dimopoulos, Dvali and Kaloper (ADDK) [18], $n$ orthogonal $(2+n)$-branes intersect on 3 spatial dimensions in $AdS_{4+n}$ spacetime, and gravity is localized to the intersection, where the Standard Model fields reside. The bulk comprises $2^n$ identical patches of $AdS_{4+n}$. In the six-dimensional theory, our world (a 3-brane) is located at the intersection of two, or more, 4-branes, so long as the tensions acting on the junction balance [19, 20]. With the Gauss-Bonnet term in the bulk action, the tension on the intersection of two orthogonal 4-branes can be nonzero [21]. Nelson's work generalizes the ADDK solution in an $AdS_6$ spacetime containing intersecting 4-branes upon which the induced metric is AdS [20]. The branes intersect on four-dimensional Minkowski junctions (3-branes) with localized gravity.

Following the observation that intersecting D-branes can give rise to chiral fermions [22], models incorporating intersecting branes in string theory vacua have been widely studied in recent years; see [23, 24] for reviews. In the intersecting D-branes scenario, the gauge fields of the Standard Model are localized on D-branes wrapping compact cycles of an internal space. The pattern of intersections governs the spectrum of the chiral fermions. Toroidal models incorporate D4, D5 or D6-branes wrapping 1, 2 or 3-cycles of $T^2$, $T^4$ (factorized as $T^2 \times T^2$) or $T^6$ (factorized as $T^2 \times T^2 \times T^2$), respectively. The D-branes wrap 1-cycles in each $T^2$. Wrapping numbers $n^i$ and $m^i$ specify the number of times each fundamental 1-cycle of the $T^2$ is wrapped. Different D-branes wrap different $(n^i, m^i)$ cycles and can intersect more than once on the torus to generate a multiplicity of particle families. Toroidal orbifold models can offer more scope for model building. In theories with two extra dimensions, $T^2/Z_2$, $T^2/Z_3$, $T^2/Z_6$ and $T^2/(Z_2 \times Z_2')$ are frequently used [13]. Supersymmetric SU(6) models have been constructed with two extra dimensions compactified on $T^2/(Z_2)^3$ and $T^2/(Z_2)^4$ [25].

In this paper, we relate particle masses to the positions of intersecting AdS 4-branes in $AdS_6$ spacetime. The branes lie upon and parallel to orbifold fixed lines and intersect on four-dimensional Minkowski junctions. In Section 2, we describe the set-up, in which 4-branes wrap 1-cycles on a rectangular $T^2/(Z_2)^3$ orbifold that is identified with $S^1/Z_2 \times S^1/(Z_2 \times Z_2')$. The radius of each $S^1$ is equal to the inverse of the AdS curvature and is of Planck scale. Four-dimensional mass parameters are related to the positions of fixed points within the $T^2/(Z_2)^3$ orbifold. In Section 3, we show that the weak gauge bosons $W^\pm$ and $Z^0$, the mesons and the charged leptons are distributed upon the orbifold fixed lines perpendicular to one of the extra dimensions, of coordinate $y$, and upon sub-lines of $2^n$-fold symmetry between the fixed lines. In Section 4, we show that the quarks are distributed upon the orbifold fixed lines perpendicular to the other extra dimension, of coordinate $z$, and that the baryons are distributed upon sub-lines of $2^n$-fold symmetry between the fixed lines. The mass differences of various hadrons in the SU(3) representations will be shown to correspond precisely to positions on orbifold fixed lines. In Section 5, we show that the quark doublets u–d, s–c and b–t are associated with fixed line and sub-line intersections. The intersections lie on the line $y = z$ in a symmetrical arrangement; in four dimensions, the corresponding sequence of fundamental scales includes the standard GUT scale of the Minimal Supersymmetric Standard Model (MSSM). We provide evidence that all particles lie on the line $y = z$, where they form symmetrical arrangements. In four dimensions, particles of all kinds occupy a geometric sequence of mass levels that descends from the Higgs field vacuum expectation value. In Section 6, we discuss our findings.

Unless stated otherwise, all values of particle mass to which reference is made have been taken from the Particle Data Group (PDG) listings for 2005 [26]. We use the values of PDG fits. Where such values are not given we use PDG averages.

Throughout the paper, for conciseness, we will refer to K$^{\pm}$ and other particle/antiparticle pairs as single particles.

**2. The model**

We work on six-dimensional Anti-de Sitter spacetime, in which 4-branes wrap 1-cycles on a rectangular orbifold T$^2$/(Z$_2$)$^3$, which is identical to S$^1$/Z$_2$ × S$^1$/(Z$_2$ × Z$_2'$) [25]. Each S$^1$ is of radius $1/k$, where $k$ is the AdS curvature. The induced metric on the 4-branes is AdS, as in [20]. The 4-branes lie parallel to the orbifold fixed lines and intersect orthogonally on 3-branes upon which the induced metric is four-dimensional Minkowski. The 3-branes are the domains of particles. The fifth dimension, of coordinate $y$, is a periodic extension of the long side of the orbifold. The sixth dimension, of coordinate $z$, is a periodic extension of the short side of the orbifold. The bulk metric is reflection symmetric about the line $y = z$. We relate mass parameters to positions in the two extra dimensions on the basis of the correspondence between scale in four-dimensional spacetime and position in a warped extra dimension [7].

First, we consider the relative sizes of the Planck scale $M$ of the higher dimensional theory, the four-dimensional Planck scale $M_P$ and the AdS curvature $k$. In the RS II model [4], with its infinite extra dimension, the scales are related by

$$M_P^2 = M^3 k^{-1}. \qquad (2)$$

To fulfil this requirement we conjecture that $M = M_P = k$. Next, we consider a scale transformation in four dimensions, from the Planck length, $1/M_P = 1/k$, to the length scale $\pi/k$ of the S$^1$/Z$_2$ orbifold. The corresponding translation in the fifth dimension is from $y = 0$, the position of the Planck brane, to $y = (1/k)\ln\pi$, the position of the fixed point across the long side of the fundamental rectangle of the T$^2$/(Z$_2$)$^3$ orbifold. In our semi-infinite extra dimension, fixed points are positioned at

$$y = (n_1/k)\ln\pi \qquad (3)$$

where $n_1$ takes the values 0, 1, 2, 3….. The warp factor $e^{-k|y|}$, by which four-dimensional mass parameters are rescaled according to position in the fifth dimension, is given by

$$e^{-k|y|} = e^{-n_1 \ln\pi} = \pi^{-n_1} \qquad (4)$$

where $n_1 \geq 0$. A geometric sequence of mass parameters in four dimensions (Planck Sequence 1) corresponds to the lattice of fixed points in the fifth dimension. The value of mass corresponding to the $n_1^{th}$ fixed point ($n_1 = 0, 1, 2, 3…..$) of the fifth dimension is given by

$$m_{n_1} = \pi^{-n_1} M_P \qquad (5)$$

where $M_P = (\hbar c/G)^{1/2}$, which has the value $1.220899 \pm 0.000090 \times 10^{19}$ GeV [27].

We now consider a scale transformation in four dimensions, from the Planck length, $1/M_P = 1/k$, to the length scale $(\pi/2)/k$ of the $S^1/(Z_2 \times Z_2')$ orbifold. The corresponding translation in the sixth dimension is from $y = 0$, the position of the Planck brane, to $y = (1/k)\ln(\pi/2)$, the position of the fixed point across the short side of the fundamental rectangle of the $T^2/(Z_2)^3$ orbifold. A (second) geometric sequence of mass parameters in four dimensions (Planck Sequence 2) corresponds to a lattice of fixed points in the semi-infinite sixth dimension. The value of mass corresponding to the $n_2^{th}$ fixed point ($n_2 = 0, 1, 2, 3\ldots$) of the sixth dimension is given by

$$m_{n_2} = (\pi/2)^{-n_2} M_P. \tag{6}$$

The numbers $n_1$ and $n_2$ refer to the wrapping of 1-cycles on the $T^2/(Z_2)^3$ orbifold by 4-branes that lie parallel to the long and short sides of the rectangle, respectively. They represent the numbers of times the orbifold is wrapped up to the positions of brane intersections. Fractional values of $n_1$ and $n_2$ parametrize fractionally wrapped branes. Fractionally wrapped branes feature in several models, including [28-30].

In Sections 3 and 4, we present values of $n_1$ and $n_2$ for the particles of the Standard Model, superimposed upon mass levels that can be viewed as graphical representations of the fixed lines that are perpendicular to the fifth and sixth dimensions.

**3. Planck Sequence 1**
We show that the mesons, charged leptons and weak gauge bosons $W^\pm$ and $Z^0$ occupy the mass levels, and sub-levels as defined below, of a geometric sequence (Planck Sequence 1), of common ratio $\pi^{-1}$, that we have identified with the positions of fixed points in the fifth dimension. Defining the $0^{th}$ term in Planck Sequence 1 as the Planck Mass $M_P$, the $n_1^{th}$ term in the sequence has a value given by (5).

The occupation of Planck Sequence 1 by the lightest charged lepton (electron), the lightest flavoured mesons ($K^\pm$ and $K^0$) and the weak gauge bosons $W^\pm$ and $Z^0$ is shown in figure 1. The electron occupies a $0^{th}$ order (integer $n_1$) level. The $K^\pm$–$K^0$ isospin doublets lie on a $0^{th}$ order mass level. $W^\pm$ and $Z^0$ form a 'mass-doublet' on a $1^{st}$ order (half-integer $n_1$) level. Mass-doublets are symmetrical configurations of particles about a mass level (fixed line or sub-line). A mass sequence of common ratio $\pi^{-3}$ appears to overlay the fundamental sequence.

Most mesons, and the muon and tau lepton, occupy 'higher order' mass levels, for which $n_1 = a.2^{-b}$ where $a$ and $b$ are positive integers; $b$ is the order of the level. The occupation of the mass levels of Planck Sequence 1 by low mass strange and charmed mesons, and states of predominantly $s\bar{s}$ and $c\bar{c}$ composition, is shown in figure 2. $K^{*\pm}$ and $K^{*0}$ occupy levels of $7^{th}$ and $8^{th}$ order, respectively. $\eta'$ occupies a $7^{th}$ order level, while $\phi$ occupies a $3^{rd}$ order level. The charmed mesons $D^0$, $D^\pm$, $D^{*0}$ and $D^{*\pm}$ are associated with two $5^{th}$ order levels. $\eta_c$ occupies a $4^{th}$ order level, while J/$\Psi$ occupies a level of $5^{th}$ order. $\Psi(2S)$ occupies a $2^{nd}$ order level.

Uncertainties in $n_1$ arising from measurement errors and from uncertainty in the Planck Mass are small compared with the level spacing. For levels of $4^{th}$ order in Planck Sequence 1, the level spacing is 0.07 of the level mass. For levels of $8^{th}$ order, the level spacing is 0.004 of the level mass.

The lowest mass unflavoured states are also found upon the mass levels of Planck Sequence 1. $\pi^0$, of mass 134.98 MeV, and $\pi^\pm$, of mass 139.57 MeV, form a mass multiplet on a $3^{rd}$ order level, of mass 137.58 MeV. $\eta$, of mass 547.75 ± 0.12 MeV, occupies an $8^{th}$ order level, of mass 547.79 MeV. ρ, of mass 775.8 ± 0.5 MeV (neutral only, e$^+$e$^-$), and ω, of mass 782.65 ± 0.12 MeV, appear to form a mass-multiplet on a 6$^{th}$ order level, of mass 779.89 MeV.

The muon, of mass 105.66 MeV, occupies an 8$^{th}$ order level in Planck Sequence 1, of mass 105.67 MeV. The tau lepton occupies a higher order level. These two charged leptons occupy low order levels in another mass sequence, as will be shown in Section 5.

**4. Planck Sequence 2**

We now provide evidence that the quarks occupy the mass levels of a geometric sequence (Planck Sequence 2), of common ratio $(\pi/2)^{-1}$, that we have identified with the positions of fixed points in the sixth dimension, and that the baryons occupy sub-levels of Planck Sequence 2. Defining the 0$^{th}$ term in Planck Sequence 2 as the Planck Mass $M_P$, the $n_2^{th}$ term in the sequence has a value given by (6).

As measures of quark mass, we use the mid-range values of the Particle Data Group's evaluations, 2003 [31], shown in table 1. The model was constructed using these evaluations. The 2005 evaluations differ slightly [26].

**Table 1.** Quark mass evaluations of the Particle Data Group, 2003 [31].

| Quark | Mass evaluation | Mid-range value used in the analysis |
|---|---|---|
| up (u) | 1.5 - 5 MeV | 3.25 MeV |
| down (d) | 5 - 9 MeV | 7 MeV |
| strange (s) | 80 - 155 MeV | 117.5 MeV |
| charm (c) | 1.0 - 1.4 GeV | 1.2 GeV |
| bottom (b) | 4.0 - 4.5 GeV | 4.25 GeV |
| top (t) | 174.3 ± 5.1 GeV; 165 ± 5 GeV ($\overline{MS}$) [32] | 165 GeV |

The up, down and strange quark masses of table 1 are $\overline{MS}$ values at a scale $\mu \approx 2$ GeV. The charm and bottom quark masses are 'running' masses in the $\overline{MS}$ scheme. The top quark mass of 174.3 ± 5.1 GeV results from direct observations of top events. The top quark running mass, $\overline{m}_t(\overline{m}_t) = 165 \pm 5$ GeV [31], has been used in the analysis for consistency with the other values of quark running mass.

Values of $n_2$ for the quark mid-range masses are presented graphically in figure 3, superimposed upon the mass levels of Planck Sequence 2. The values of quark mass lie close to mass levels (fixed lines). The level numbers, $n_2$, corresponding to the quark mass evaluations of the Particle Data Group [31] are presented in table 2. Included in the table are the quark masses of the model.

**Table 2.** Quark masses of the model.

| Quark | Level number $n_2$, from (6), corresponding to PDG mid-range mass | Quark mass of the model |
|---|---|---|
| u | 110.0 | 3.262 MeV ($n_2 = 110$) |
| d | 108.3 | 8.049 MeV ($n_2 = 108$) |
| s | 102.1 | 120.9 MeV ($n_2 = 102$) |
| c | 96.9 | 1.156 GeV ($n_2 = 97$) |
| b | 94.1 | 4.481 GeV ($n_2 = 94$) |
| t | 86.0 | 166.1 GeV ($n_2 = 86$) |

The baryons are found to lie upon higher order mass levels in Planck Sequence 2. Here, we show the occupation of mass levels by isospin singlet states of each flavour and by low mass unflavoured and flavoured baryons. Values of $n_2$ for the $I(J^P) = 0(½^+)$ baryons, $\Lambda$ (uds), $\Lambda_c^+$ (udc), $\Lambda_b^0$ (udb) and $\Omega_c^0$ (ssc), are presented in figure 4. $\Lambda_c^+$ and $\Lambda_b^0$ appear to occupy 1$^{st}$ order levels, while $\Omega_c^0$ occupies a 3$^{rd}$ order level. $\Lambda$ (uds, $I$=0) forms a symmetrical mass-doublet with $\Sigma^0$ (uds, $I$=1), centred on Level 97. The mass difference, 76.959 ± 0.023 MeV, of these two strange baryons corresponds to a value of $n_2$, in (5), of 103.001 ± 0.003. In other words, the mass difference is equal to the mass of a 0$^{th}$ order level. Precise 0$^{th}$ order mass differences also occur between $\Sigma$ baryons and other $|S|$=1 light hadrons, as will be shown later in this section. The lowest mass unflavoured, strange and charmed baryons are associated with 7$^{th}$ and lower order mass levels. Values of $n_2$ for these baryons are presented in figure 5. The proton-neutron isospin doublet forms a mass-doublet on a 7$^{th}$ order level. The $\Delta(1232)$ states are centred upon a 6$^{th}$ order level, of mass 1232.1 MeV. We saw in figure 4 that $\Lambda$ and $\Sigma^0$ form a mass-doublet on a 0$^{th}$ order level; in figure 5, $\Sigma^+$ and $\Sigma^-$ are shown to occupy 6$^{th}$ order levels. The singlet state, $\Lambda_c^+$ lies close to a 1$^{st}$ order level, as was shown in figure 4, but occupies a 7$^{th}$ order level.

Uncertainties in $n_2$ arising from measurement errors and from uncertainty in the Planck Mass (0.007%) are small compared with the level spacing. For levels of 3$^{rd}$ order in Planck Sequence 2, the level spacing is 0.06 of the level mass; for levels of 7$^{th}$ order, the level spacing is 0.004 of the level mass.

Values of mass difference between $\Sigma$ baryons and other $|S|$=1 hadrons of the SU(3) multiplets equal the masses of 0$^{th}$ order levels in Planck Sequence 2. As we have stated, the mass difference of the uds baryons, $\Lambda$ ($I = 0$) and $\Sigma^0$ ($I = 1$), equals the mass of a 0$^{th}$ order level ($n_2 = 103$) in Planck Sequence 2. The mass difference of $\Sigma^0$ and $\Sigma(1385)^0$ is 191.1 ± 1.0 MeV, for which $n_2 = 100.99 \pm 0.01$. $\Sigma^+$ (uus) and $\Sigma^-$ (dds) differ in mass by 8.08 ± 0.08 MeV, for which $n_2 = 107.99 \pm 0.02$. The value of $n_2$ corresponding to the mass difference of K*$^\pm$ ($u\bar{s}$) and $\Sigma^0$ (uds) equals 99.98, while the value of $n_2$ corresponding to the mass difference of K*$^0$ ($d\bar{s}$) and $\Sigma^0$ equals 100.01. The quark content of K*$^\pm$ is given as that of the positively charged meson. The two values of mass difference closely straddle a 0$^{th}$ order level. Values of $n_2$ corresponding to strange hadron mass differences, and to the mass of the strange quark, are presented in figure 6. Those mass differences featuring strange hadrons differing in isospin or spin, or both, are described below, in terms of the strange quark mass $m_s$ of the model:

$$m_{\Sigma^0} - m_\Lambda = (\pi/2)^{-1} m_s \qquad (7)$$

$$m_{\Sigma(1385)^0} - m_{\Sigma^0} = (\pi/2) m_s \qquad (8)$$

$$m_{\Sigma^0} - m_{K^*} = (\pi/2)^2 m_s \qquad (8)$$

The mass difference of $\Sigma^+$ and $\Sigma^-$ equals the mass of the down quark of the model:

$$m_{\Sigma^-} - m_{\Sigma^+} = m_d \qquad (10)$$

The mass difference of the $\phi$ resonance ($s\bar{s}$), of hidden flavour, and the $\Xi^0$ (uss) - $\Xi^-$ (dss) isospin doublet also equals the mass of a 0$^{th}$ order level. That is, the value of $n_2$ corresponding to the mass difference of $\phi$ and $\Xi^0$ equals 100.02, while the value of $n_2$ corresponding to the mass difference of $\phi$ and $\Xi^-$ equals 99.97. The two values closely straddle Level 100, as did the values of $n_2$ corresponding to the mass differences of K*$^\pm$ and $\Sigma^0$, and K*$^0$ and $\Sigma^0$. The mass difference of $\phi$ and $\Xi$ is given by

$$m_\Xi - m_\phi = (\pi/2)^2 m_s \qquad (11)$$

Equations (7) - (11) describe relationships that are precise to within the small experimental uncertainties.

**5. On the line y = z**
We show that scales identified with the three families of the Standard Model correspond to three positions within the extra-dimensional space where low order fixed lines of the $T^2/(Z_2)^3$ orbifold coincide to high precision at $y = z$. Then we provide evidence that all particles occupy positions on the line $y = z$.

The u–d quark mass-doublet is centred on a mass level in Planck Sequence 2 for which $n_2 = 109$ (see figure 3). The value of $n_1$ corresponding to the mass of this level is 42.9992. For all values of mass,

$$n_2 n_1^{-1} = \ln\pi [\ln(\pi/2)]^{-1} \qquad (12)$$

which has the value 2.53493. The quotient of the two prime numbers 109 and 43, the values of $n_2$ and $n_1$ associated with the u-d mass-doublet, is 2.53488. The fraction 109/43 is closer in value to $\ln\pi[\ln(\pi/2)]^{-1}$ than any other fraction constructed from integers in the range 1 to 1000, other than those formed from multiples of 109 and 43. Since the fraction 38/15 (2.53333) is also close in value to $\ln\pi[\ln(\pi/2)]^{-1}$, low (0$^{th}$ to 2$^{nd}$) order mass levels in Planck Sequence 1 and Planck Sequence 2 coincide on the line $y = z$ where $n_1 = 43$ and $n_2 = 109$, $n_1 = 43 - 15/4 = 39.25$ and $n_2 = 109 - 38/4 = 99.5$, and $n_1 = 43 - 15/2 = 35.5$ and $n_2 = 109 - 38/2 = 90$. In Planck Sequence 2, the quark mass-doublets are centred on levels (fixed lines) for which $n_2$ takes the values 109, 99.5 and 90 (see figure 3). We conclude that the three quark mass-doublets, and thereby families of the Standard Model, are associated with three positions on the line $y = z$, up to which orthogonal 1-cycles of the $T^2/(Z_2)^3$ orbifold are either fully wrapped or fractionally wrapped, the fraction being a half-integer or quarter-integer.

Various prominent hadrons and hadron mass-doublets are also associated with low order levels in Planck Sequence 1 and Planck Sequence 2, suggesting that they also lie on the line $y = z$. Consider the strange hadron pairs, K$^\pm$-K$^0$ and $\Lambda$-$\Sigma^0$ of the SU(3) representations. The meson pair K$^\pm$-K$^0$ is associated with a 0$^{th}$ order level in Planck Sequence 1, as shown in figure 1, and a 3$^{rd}$ order level in Planck Sequence 2. The baryon pair $\Lambda$-$\Sigma^0$ forms a mass-doublet on a 0$^{th}$ order level in Planck Sequence 2, as shown in figure 4, and a 3$^{rd}$ order level in Planck Sequence 1. The weak gauge boson pair W$^\pm$–Z$^0$ is also associated with low order levels in both Planck sequences. The two gauge bosons were shown in figure 1 to form a mass-doublet on a 1$^{st}$ order level in Planck Sequence 1. They also form a mass-doublet on a 1$^{st}$ order level in Planck Sequence 2. We can also identify singlet states that occupy low order levels in both Planck Sequence 1 and Planck Sequence 2. The baryon $\Lambda_b^0$ occupies a 1$^{st}$ order level in Planck Sequence 2, as shown in figure 4, and a 3$^{rd}$ order level in Planck Sequence 1. The vector meson resonance $\phi$ occupies a 3$^{rd}$ order level in Planck Sequence 1, as shown in figure 2, and a 5$^{th}$ order level in Planck Sequence 2. Many other particles appear to occupy higher order levels in both Planck Sequence 1 and Planck Sequence 2, suggesting that all particles lie on the line $y = z$.

We have seen that the lowest mass particles with specific quantum numbers, *e.g.* $I(J^P) = 0(\frac{1}{2}^+)$, occupy low order levels in their respective sequences. Now we see, in figure 7, that on the line $y = z$, the lowest mass baryon (proton), the lowest mass charged meson ($\pi^\pm$), the lowest mass charged lepton (electron) and the quark pairs form a symmetrical arrangement that incorporates the Higgs field vacuum expectation value (VEV). Each quark pair is represented in figure 7 by the wrapping numbers ($n_1$, $n_2$) associated with the mass-doublet. In the Standard Model, the quark and lepton masses are generated through couplings with the Higgs field. The mass of fermion $f$ is given by $m_f = c_f v$, where $c_f$ is a dimensionless coupling constant and $v = 2^{-1/4} G_F^{-1/2}$ is the Higgs field VEV, equal to 246.22 GeV [27]. We now show that, on the line $y = z$, particles in general (weak gauge bosons, quarks, hadrons and leptons) occupy positions related to that corresponding to the Higgs field VEV. The corresponding 'Higgs Sequence' of particle masses descends from the Higgs field VEV with common ratio $r_H = (\pi/2)^{-5}$ and contains its own higher order levels. Level numbers in the Higgs Sequence are denoted by $n_H$. The occupation of the mass levels of the Higgs Sequence by the vector boson $W^\pm$, the strange and charm quarks of Planck Sequence 2, the lowest mass strange mesons and baryons, and the lowest mass mesons and baryons constructed from strange and charm quarks, is shown in figure 8. Of the hadrons, strange, 'charmed, strange' and neutral vector states occupy the lowest order levels. The occupation of the Higgs Sequence by neutral vector states is considered separately. $W^\pm$ occupies a 1$^{st}$ order mass level ($n_H = 0.5$) in the Higgs Sequence. The strange and charm quarks of Planck Sequence 2 occupy 3$^{rd}$ order mass levels. The lightest strange meson, $K^\pm$, forms a mass-doublet with its isospin partner, $K^0$, on a 2$^{nd}$ order level. The lightest charmed, strange meson, $D_s^\pm$ ($J=0$), forms a mass-doublet with its spin partner, $D_s^{*\pm}$ ($J=1$), on a 3$^{rd}$ order level, the two mesons occupying mass levels of 6$^{th}$ order. The lightest strange baryon, $\Lambda$ (uds, $I=0$), forms a mass-doublet with $\Sigma^0$ (uds, $I=1$) on a 3$^{rd}$ order level, the two baryons occupying mass levels of 6$^{th}$ order. The baryon $\Omega_c^0$ (ssc) occupies a 0$^{th}$ order mass level. Finally, the elusive exotic baryon $\theta^+$ ($uudd\bar{s}$) occupies a 2$^{nd}$ order level in the Higgs Sequence. $\theta^+$ occupies a third order level in Planck Sequence 2. The values of $n_H$ and $n_2$ associated with $\theta^+$ are $2.249 \pm 0.001$ and $96.374 \pm 0.003$, respectively.

The neutral vector boson, $Z^0$, and the unflavoured neutral vector meson resonances, J/$\Psi$ ($c\bar{c}$) and Y ($b\bar{b}$) occupy mass levels of 4$^{th}$ order in the Higgs Sequence, as shown in figure 9. The neutral vector meson, $B_s^{*0}$ ($s\bar{b}$) occupies a level of 4$^{th}$ order between the levels occupied by J/$\Psi$ and Y. The SU(3) octet-singlet mixed neutral vector states, $\omega$ and $\phi$, occupy higher order levels as shown. Relative to the mass of the $Z^0$ boson, $\phi$, J/$\Psi$ and Y occupy 0$^{th}$ and 1$^{st}$ order levels in a geometric sequence, the $Z^0$ Sequence, of common ratio $(\pi/2)^{-5}$, as shown in figure 10. In this arrangement, $B_s^{*0}$ occupies a level of 2$^{nd}$ order. The tau lepton is included in the $Z^0$ Sequence since it supplements the symmetry in level occupation, occupying a level of 2$^{nd}$ order between the levels occupied by $\phi$ and J/$\Psi$. Within the small experimental uncertainty, the mass of the tau lepton (1776.99 +0.29/-0.26 MeV) equals the geometric mean of the masses of $\phi$ and J/$\Psi$ (1776.85 $\pm$ 0.03 MeV):

$$m_\tau = (m_\phi m_{J/\Psi})^{1/2} \quad (13)$$

The muon occupies a 0$^{th}$ order level within the $Z^0$ Sequence and, like the tau lepton, supplements the symmetry in vector meson level occupation. The masses of the two leptons are related through the equation

$$m_\tau m_\mu^{-1} = (\pi/2)^{25/4} \quad (14)$$

to within the small experimental uncertainty. The presence of charged leptons in a vector sequence presumably reflects their nature in the higher dimensional theory.

There is also evidence for a pseudoscalar sequence within the Higgs Sequence. The unflavoured neutral pseudoscalar mesons of the SU(4) 20-plet occupy the sub-levels of an *ad hoc* bottom-up sequence defined by the mass of the lightest pseudoscalar meson, $\pi^0$. Relative to the mass of $\pi^0$, the states $\eta$, $\eta'$ and $\eta_c$ occupy levels of 3$^{rd}$ order in a geometric sequence of common ratio $(\pi/2)^5$, as shown in figure 11. The neutral pseudoscalar meson, $B_s^0$ ($s\bar{b}$) also occupies a 3$^{rd}$ order level in the arrangement. The Particle Data Group [25] notes one observation of $\eta_b$, of mass 9300 ± 20 ± 20 MeV, although this observation needs to be confirmed. The corresponding value of $[n_H - n_H(\pi^0)]$ is −1.875 ± 0.002. In other words, $\eta_b$ might also occupy a 3$^{rd}$ order level in the pseudoscalar sequence.

Finally, we touch on the question of gauge coupling unification in an intersecting brane world. Blumenhagen, Lust and Stieberger [33] have shown that three generation supersymmetric intersecting brane world models can lead to gauge coupling unification at a scale close to the standard GUT scale of the Minimal Supersymmetric Standard Model, $2 \times 10^{16}$ GeV. In figure 7, we saw that the three families of the Standard Model are associated with three 4-brane intersections on the line $y = z$, at scales related by powers of a factor $(\pi/2)^{9.5}$. This sequence of fundamental scales includes the GUT scale. The scale associated with the wrapping numbers (5.5, 14) is $2.2 \times 10^{16}$ GeV.

## 6. Discussion

The matter content of our world arises at 4-brane intersections in AdS$_6$ spacetime. The mass of a particle in four dimensions depends on its distance from the Planck brane in either extra dimension, since particles are situated upon a line equidistant from the two extra dimensions. We can relate the same phenomena to the geometry of both AdS$_6$ and AdS$_5$ spacetimes. In AdS$_6$, 4-branes wrap 1-cycles of the orbifold $T^2/(Z_2)^3 = S^1/Z_2 \times S^1/(Z_2 \times Z_2')$, and intersect on 3-branes, which are the domains of particles. The scales of the Standard Model families relate to the GUT scale of the MSSM and derive from the geometry of the $T^2/(Z_2)^3$ orbifold. In AdS$_5$, 3-branes occupy positions in an extra dimension that is compactified on the orbifold $S^1/(Z_2 \times Z_2')$. In the latter model, the mass of a particle in four dimensions depends on its distance from the Higgs brane in the sixth dimension.

We have shown that the values of couplings with the Higgs field have a geometrical interpretation within the model. However, the value of the Higgs Sequence common ratio, $(\pi/2)^{-5}$, remains a puzzle. Five wrappings of a 1-cycle parallel to the sixth dimension correspond to 1.97 wrappings of a 1-cycle parallel to the fifth dimension, so perhaps the value $(\pi/2)^{-5}$ results from almost coincidental integer wrappings of the $T^2/(Z_2)^3$ orbifold.

Some hadron mass differences appear to depend on differences in the quantum numbers of the related particles. Equations (7-11) show that some mass differences can be related to differences in *I* or *J*.

The particles participating in a mass-doublet, in the Planck sequences and the Higgs Sequence, differ from each other with respect to charge, spin, isospin or isospin projection. Some mass-doublets might arise as a consequence of D-brane splitting. The splitting of branes that wrap parallel cycles can occur in toroidal orbifolds because of the non-rigid nature of the cycles [34].

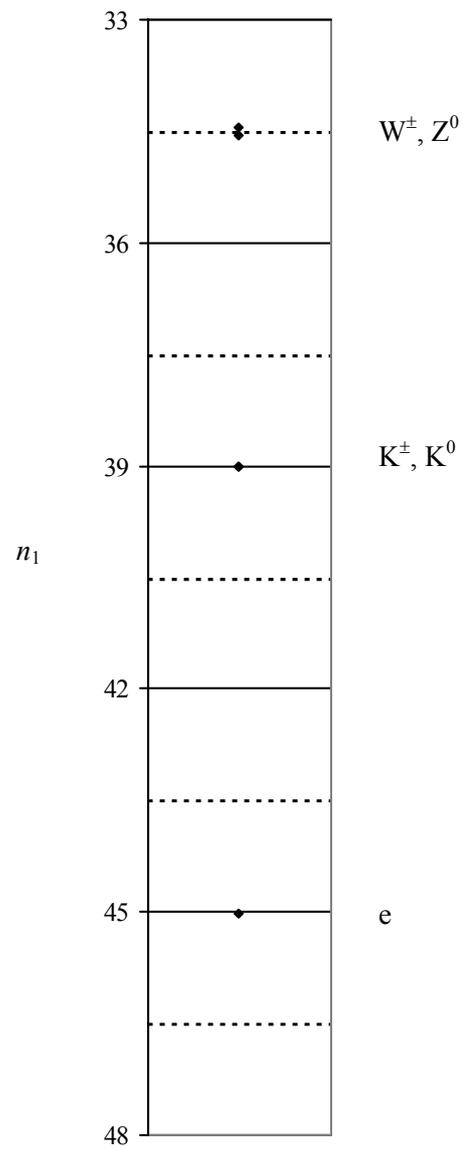

**Figure 1.** The distribution in Planck Sequence 1 of the lightest charged lepton (electron), the lightest flavoured hadrons (K-mesons) and the weak gauge bosons, $W^{\pm}$ and $Z^0$.

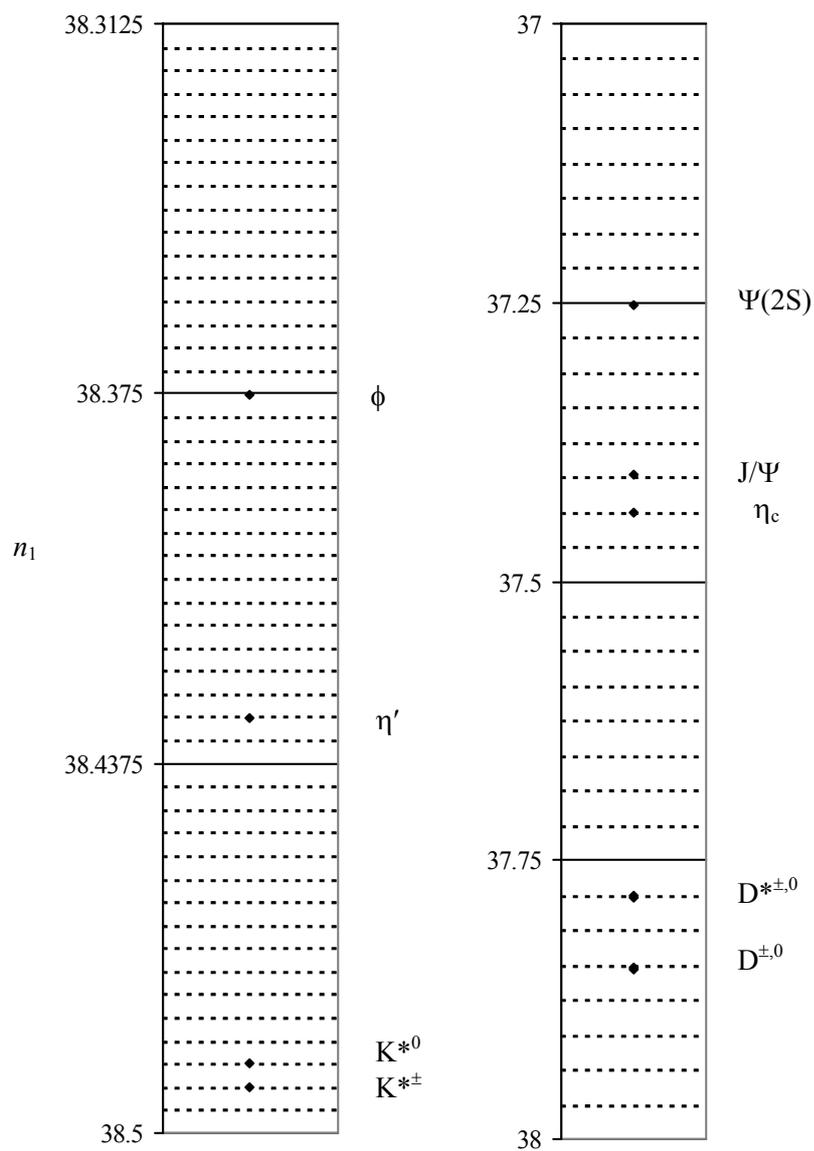

**Figure 2.** The distribution in Planck Sequence 1 of low mass strange and charmed mesons, and $s\bar{s}$ and $c\bar{c}$ states. The sub-levels shown are of 7$^{th}$ and 8$^{th}$ order.

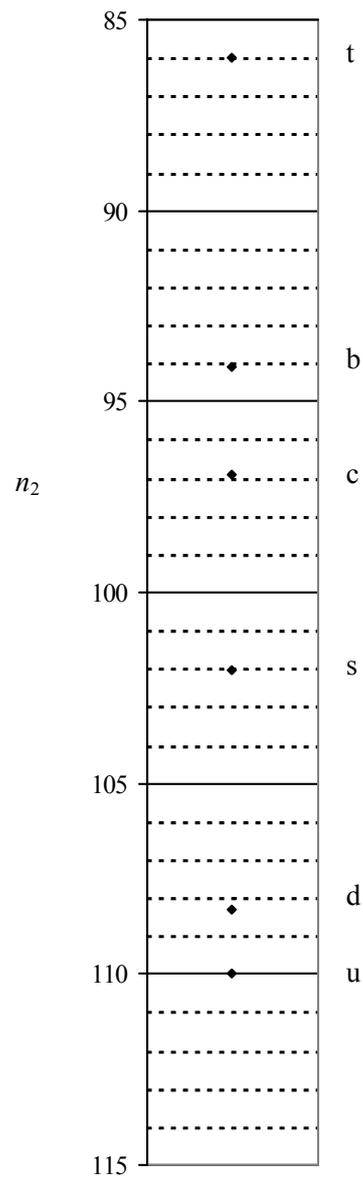

**Figure 3.** The distribution in Planck Sequence 2 of the quarks. As measures of quark mass, we use the mid-range values of the Particle Data Group's evaluations, 2003 [31].

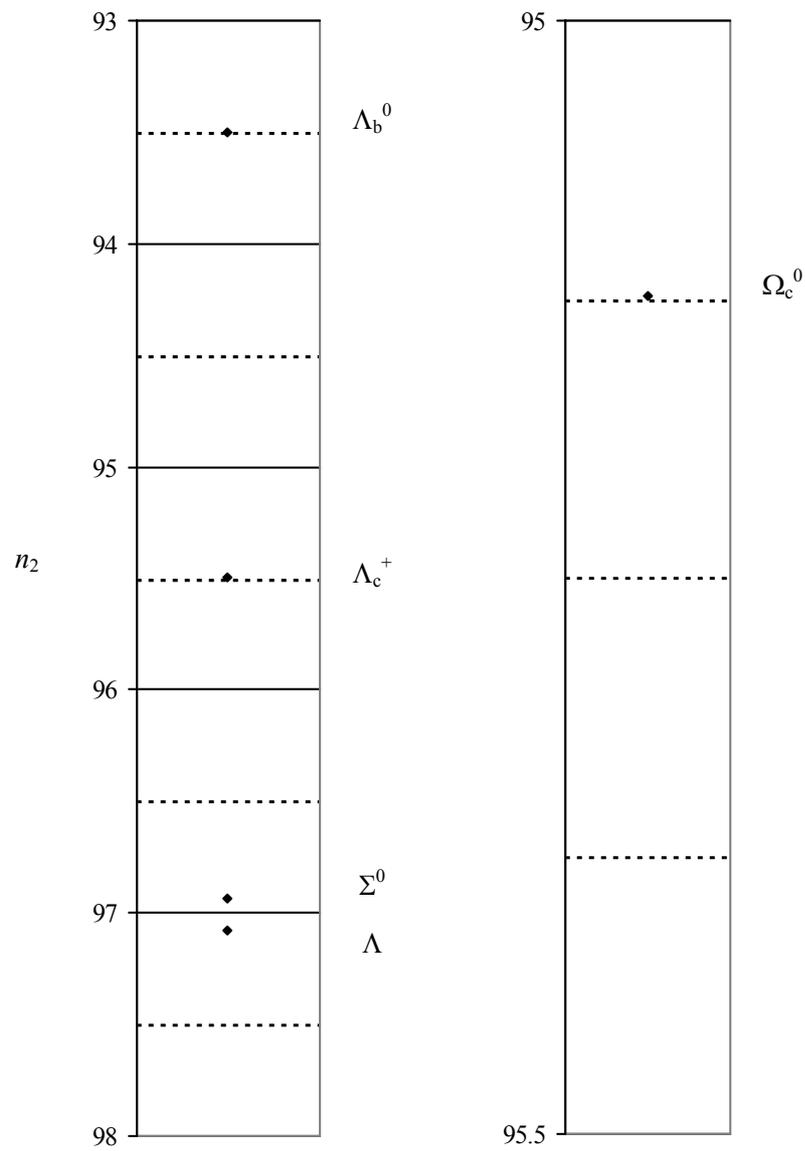

**Figure 4.** The distribution in Planck Sequence 2 of $I(J^P) = 0(½^+)$ baryons. $\Sigma^0$ is included as it forms a 'mass-doublet' with $\Lambda$.

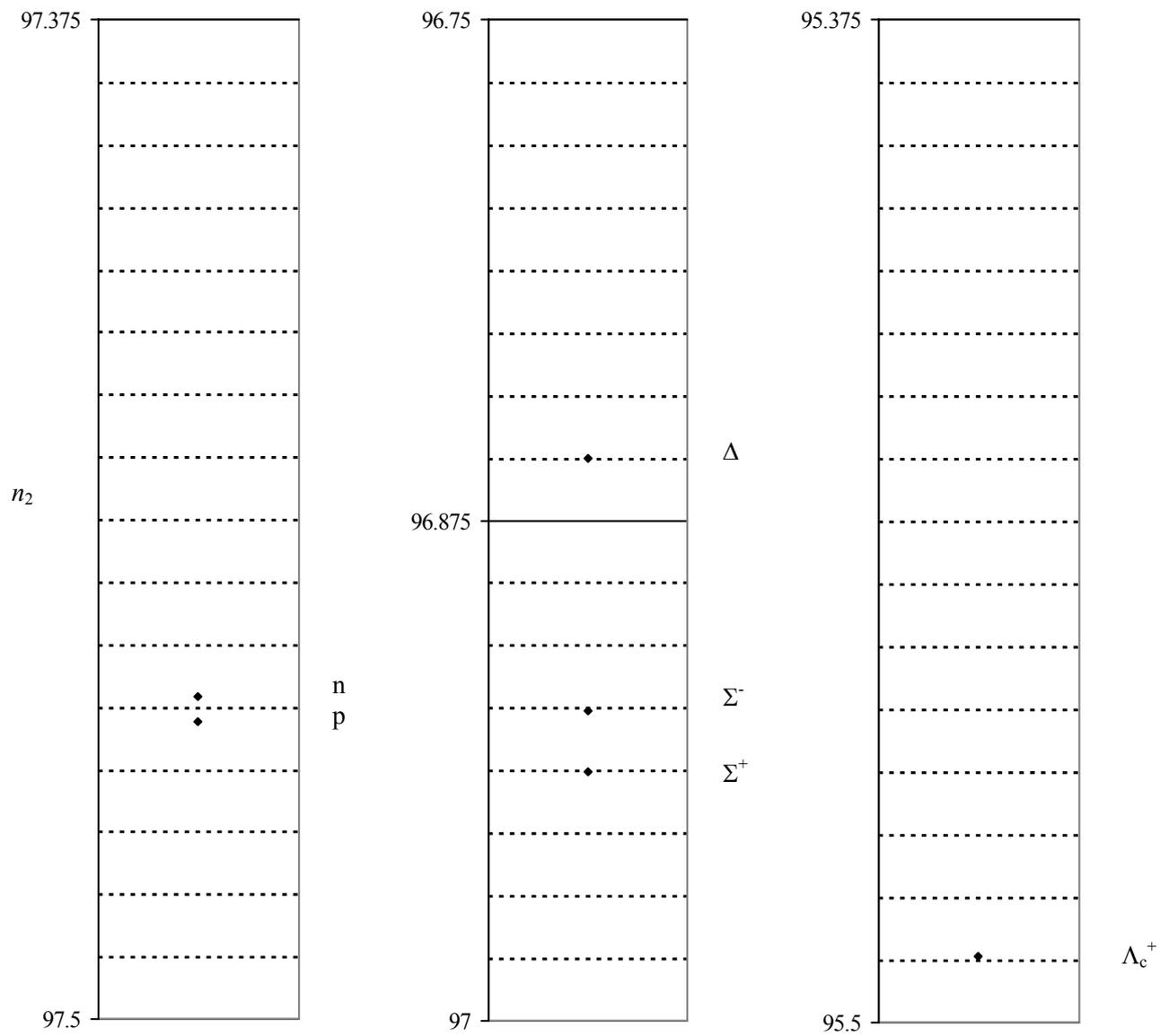

**Figure 5.** The distribution in Planck Sequence 2 of low mass unflavoured, strange and charmed baryons. The sub-levels shown are of 6$^{th}$ and 7$^{th}$ order.

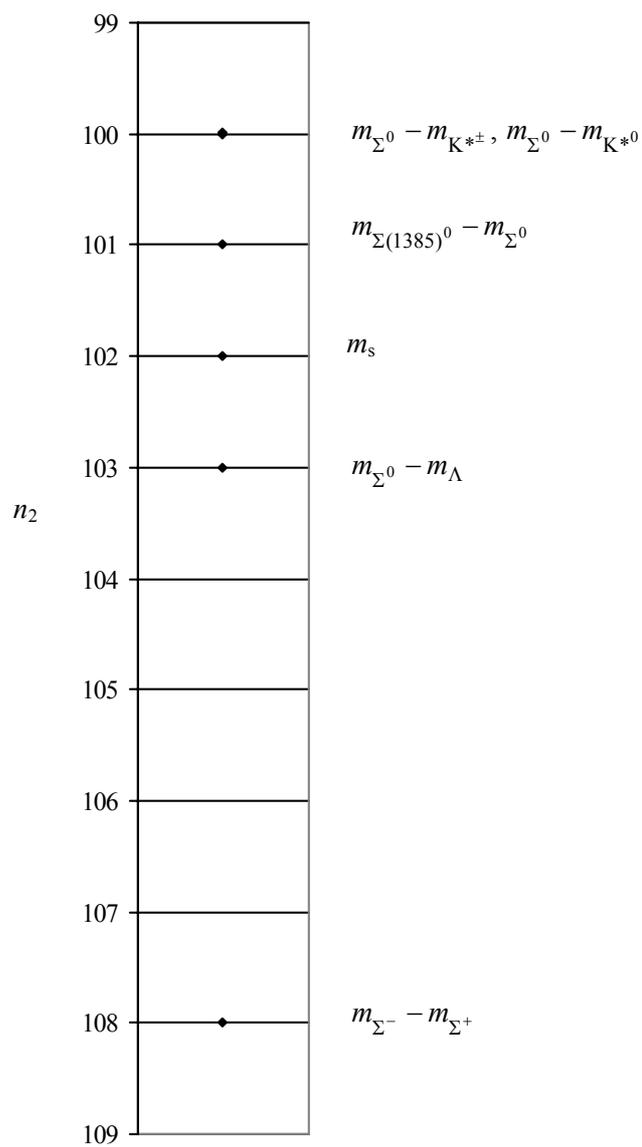

**Figure 6.** The distribution in Planck Sequence 2 of strange hadron mass differences.

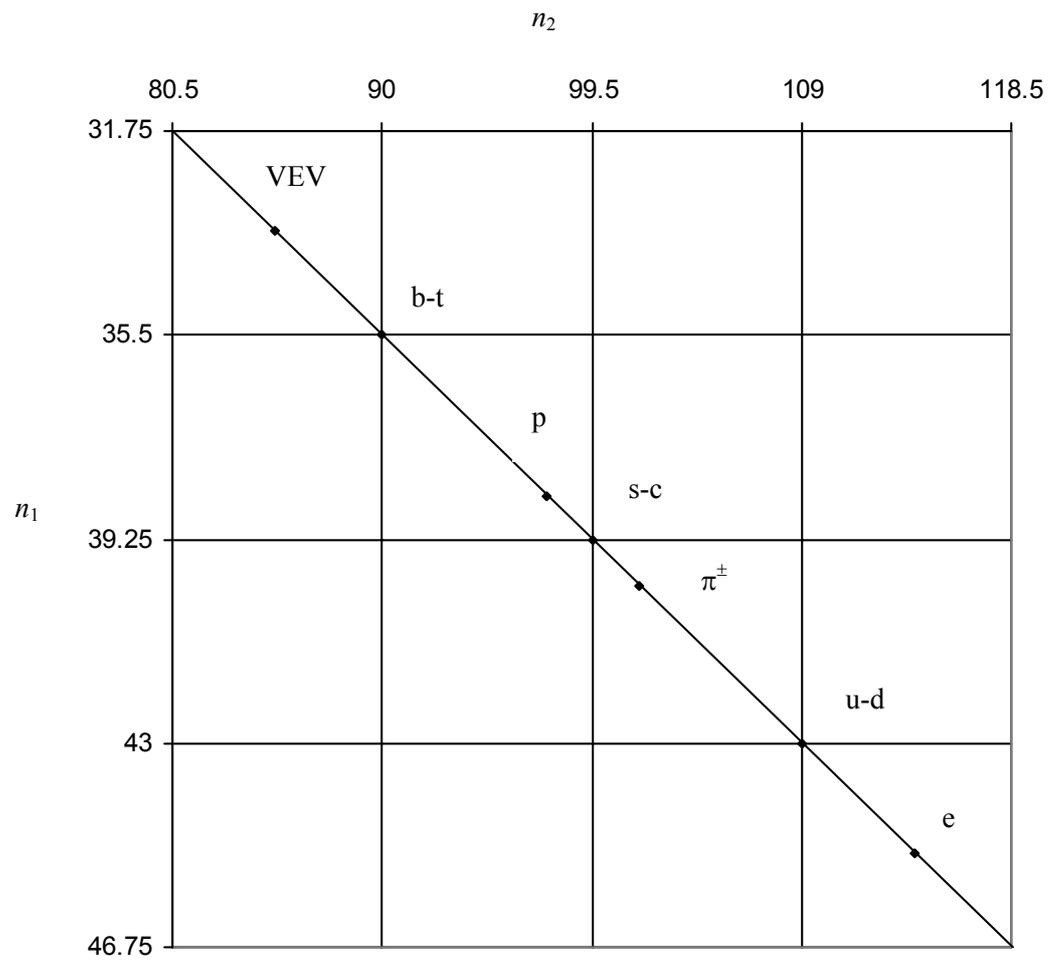

**Figure 7.** On the line y = z.

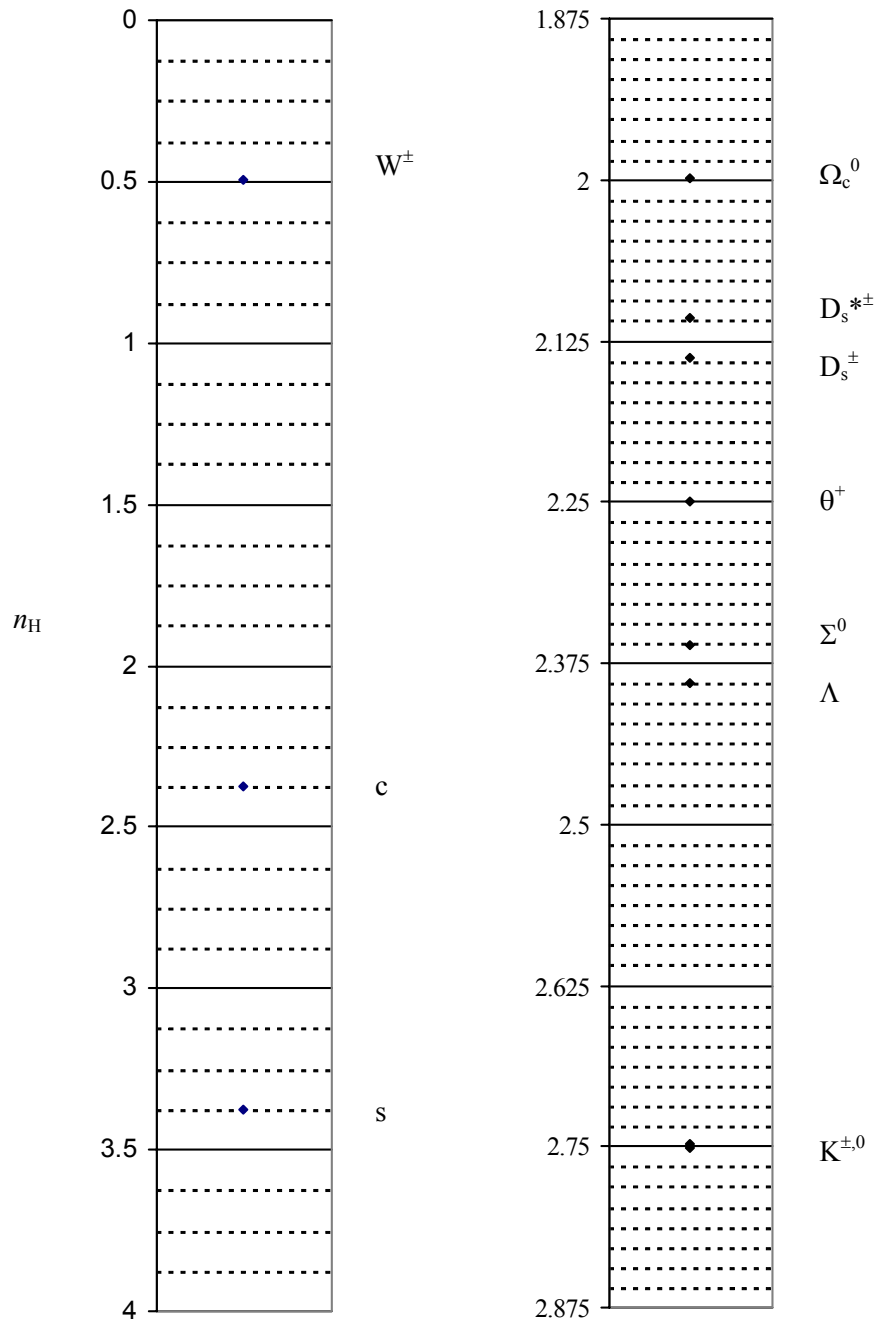

**Figure 8.** The distribution in the Higgs Sequence of the weak gauge boson $W^{\pm}$, the strange and charm quarks of the model, the lowest mass strange mesons and baryons, and the lowest mass mesons and baryons constructed from strange and charm quarks.

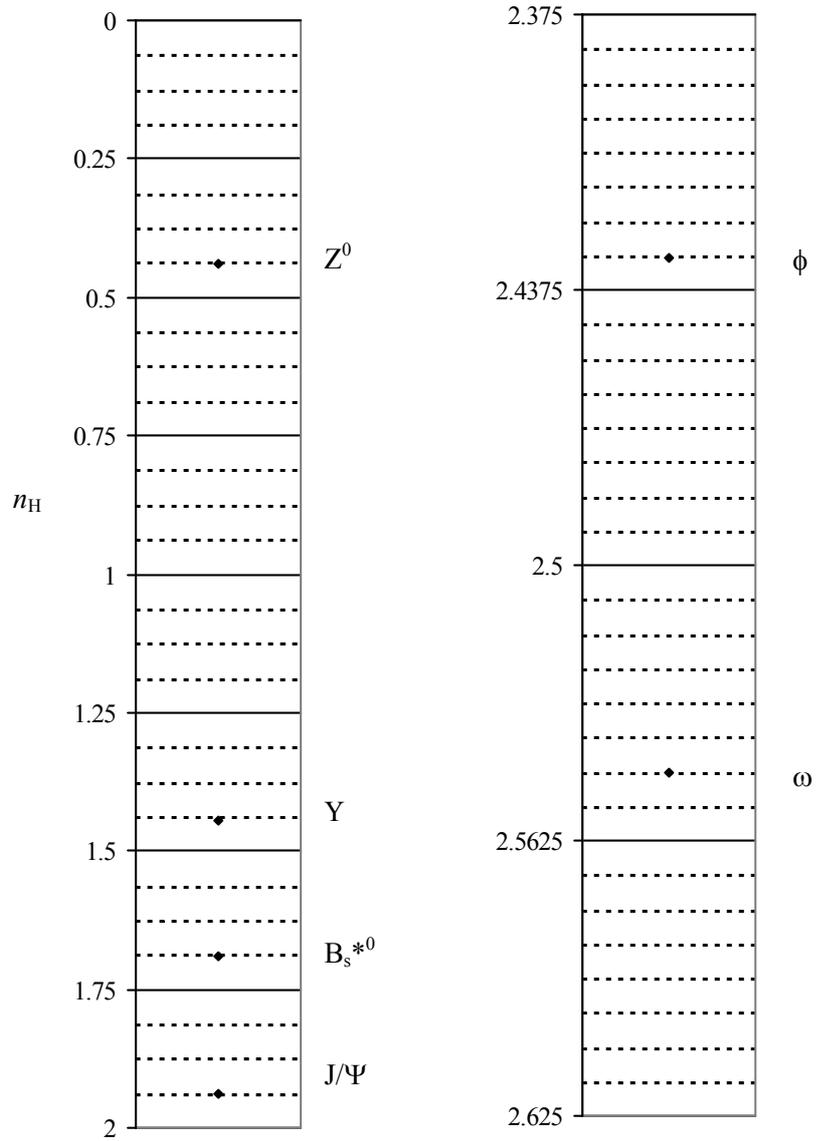

**Figure 9.** The distribution in the Higgs Sequence of the weak gauge boson $Z^0$ and neutral vector singlet states.

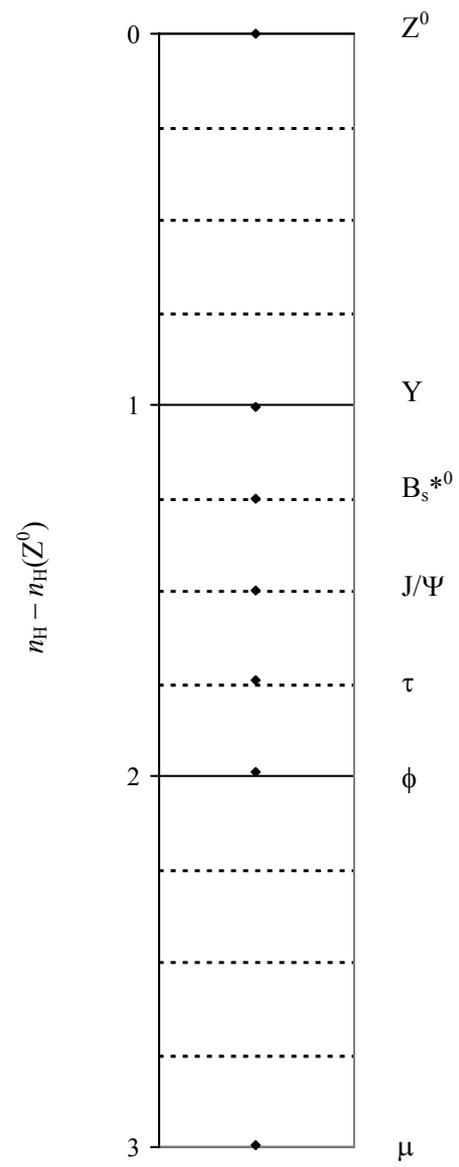

**Figure 10.** The $Z^0$ Sequence.

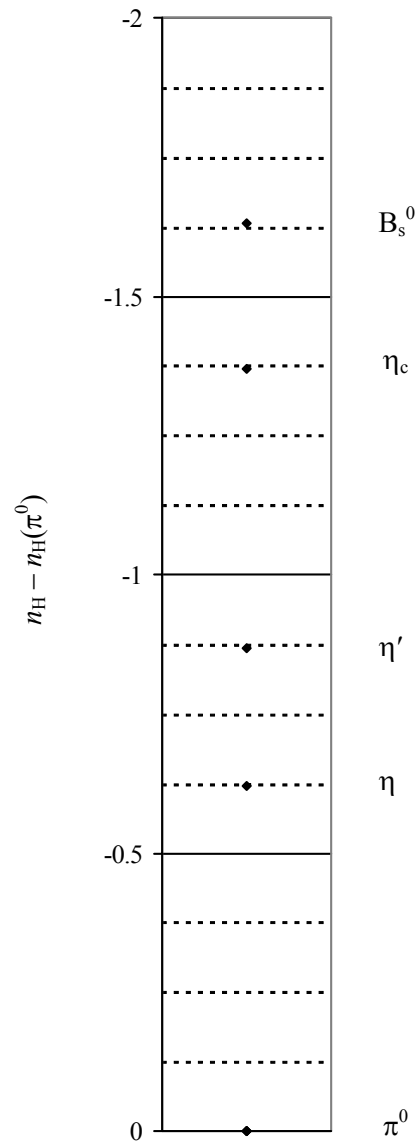

**Figure 11.** An *ad hoc* pseudoscalar sequence.